\documentclass[sigconf]{acmart}

\AtBeginDocument{%
  \providecommand\BibTeX{{%
    \normalfont B\kern-0.5em{\scshape i\kern-0.25em b}\kern-0.8em\TeX}}}

\copyrightyear{2026}
\acmYear{2026}
\setcopyright{cc}
\setcctype{by}
\acmConference[CHI '26]{Proceedings of the 2026 CHI Conference on Human Factors in Computing Systems}{April 13--17, 2026}{Barcelona, Spain}
\acmBooktitle{Proceedings of the 2026 CHI Conference on Human Factors in Computing Systems (CHI '26), April 13--17, 2026, Barcelona, Spain}
\acmPrice{}
\acmDOI{10.1145/3772318.3790872}
\acmISBN{979-8-4007-2278-3/2026/04}

\usepackage{multirow,acmart-taps} %

\usepackage{bm}       
\usepackage{amsfonts}

\usepackage{graphicx}
\usepackage{subcaption}

\usepackage{fixltx2e}
\usepackage{float}

\usepackage{makecell}

\def\thickhline{\noalign{\hrule height1pt}}

\newlength\replength
\newcommand\repfrac{.33}

\newcommand\rulewidth{.6pt}
\newcommand\tdashfill[1][\repfrac]{\cleaders\hbox to \replength{%
  \smash{\rule[\arraystretch\ht\strutbox]{\repfrac\replength}{\rulewidth}}}\hfill}

\newcommand\tdotfill[1][\repfrac]{\cleaders\hbox to \replength{%
  \smash{\raisebox{\arraystretch\dimexpr\ht\strutbox-.1ex\relax}{.}}}\hfill}

\newcommand\sub[1]{\textsubscript{#1}}

\newcounter{casecounter}[section]
\renewcommand{\thecasecounter}{\arabic{casecounter}}

\newcommand{\case}[1]{
  \refstepcounter{casecounter}
  \thecasecounter
}

\begin{document}

%
%
%
%
%

\title{Accessibility-Driven Information Transformations in Mixed-Visual Ability Work Teams}

\author{Yichun Zhao}
\orcid{0009-0006-8239-6011}
\affiliation{
\department{Department of Computer Science}\institution{University of Victoria}
\city{Victoria}
\state{British Columbia}
\country{Canada}}
\email{yichunzhao@uvic.ca}

\author{Miguel A. Nacenta}
\orcid{0000-0002-9864-9654}
\affiliation{
\department{Department of Computer Science}\institution{University of Victoria}
\city{Victoria}
\state{British Columbia}
\country{Canada}}
\email{nacenta@uvic.ca}

\author{Mahadeo A. Sukhai}
\orcid{0000-0002-9122-2519}
\affiliation{\institution{IDEA-STEM}
\city{Kingston}
\state{Ontario}
\country{Canada}}
\email{mahadeo.sukhai@idea-stem.ca}

\author{Sowmya Somanath}
\orcid{0009-0005-8580-5215}
\affiliation{
\department{Department of Computer Science}\institution{University of Victoria}
\city{Victoria}
\state{British Columbia}
\country{Canada}}
\email{sowmyasomanath@uvic.ca} 
\begin{abstract}
    %


Blind and low-vision (BLV) employees in mixed-visual ability teams often encounter information (e.g., PDFs, diagrams) in
inaccessible formats. To enable teamwork, teams must transform these representations by modifying or re-creating them into accessible forms. However, these transformations are frequently overlooked, lack infrastructural support, and cause additional labour. To design systems that move beyond one-off accommodations to effective mixed-ability collaboration, we need a deeper understanding of the
representations, their transformations and how they occur. We conducted a week-long diary study with follow-up interviews
with 23 BLV and sighted professionals from five legal, non-profit, and consulting teams, documenting 36 transformation cases.
Our analysis characterizes how teams perform representational transformations for accessibility: how they
are triggered proactively or reactively, how they simplify or enhance, and four common patterns in which workers coordinate
with each other to address representational incompatibility. Our findings uncover opportunities for designing systems that can better
support mixed-visual ability work.

 \end{abstract}



\maketitle

\section{Introduction}
Teams of knowledge workers rely heavily on shared representations such as documents, visual diagrams and spreadsheets to externalize thoughts, organize data, and facilitate coordination~\cite{Hutchins_1995, Kirsh_2010, devries-LearningExternalRepresentations-2012, Davis_Shrobe_Szolovits_1993, zhangRepresentationsDistributedCognitive1994}, all necessary towards accomplishing their goals. When the teams are perceptually diverse, one of the impediments to teamwork is that shared representations are not accessible in the same way across individuals in the team, making information difficult or impossible to access for members who need it to carry out their tasks or, more generally, keep awareness of the work of others. Teams and individuals must then find ways to cope with the situation, 
typically resulting in inefficient ad-hoc workarounds~\cite{Marathe_Piper_2025, Mörike_Kiossis_2024} and \emph{invisible labour}~\cite{Branham_Kane_2015, Cha_DIYDilemma} that can place inequitable additional burdens on some members of the team.

More specifically, when information contained in a representation is not accessible, making it available requires a \emph{representational transformation}. Representational transformations (e.g., turning a diagram into a paragraph of text or \emph{vice versa}) are a useful part of knowledge work (e.g.,~\cite{binksRepresentationalTransformationsUsing2022,Bobek_Tversky_2016,Eilam_Ofer_2018}), but they take additional effort. When set off by a lack of accessibility, the process of transforming representations can slow down work and might contribute to inequity in the distribution of work in teams.

Although this type of invisible labour has been touched upon in previous analyses of mixed-ability teams (in work activities~\cite{Branham_Kane_2015, Potluri_Pandey_Begel_Barnett_Reitherman_2022, Das_IdeationDisabledProfessionals, Kaschnig_Neumayr_Augstein_2024, Mack_Das_Jain_Bragg_Tang_Begel_Beneteau_Davis_Glasser_Park_etal._2021} and other contexts~\cite{Xiao_Bandukda_Angerbauer_Lin_Bhatnagar_Sedlmair_Holloway_2024}), there has not been an examination of mixed ability teams that considers specifically how representational barriers impede work, and how different teams engage in representational transformation to address these barriers. We believe that a better understanding of this aspect of knowledge work is worthy of examination because of two main reasons: a) we suspect that it is pervasive and; b) it offers opportunities for improving software for group work, since representations, how they are managed, integrated and transformed is a major part of the design of information systems.

To investigate mixed-visual ability teamwork from the lens of representations and representational transformations, we carried out an in-the-wild qualitative study of BLV workers and their sighted colleagues. 
We conducted a five-day diary study and follow-up interviews with 23 workers from five mixed-visual teams, and two separate focus groups with two other teams. 
The study seeks to shed light on how teams carry out accessibility-driven transformations and on the consequences of these processes.

Our data indeed contained numerous instances of representations with accessibility issues and the labour of representational transformation, which offers some justification for our motivation. Through the data analysis we arrived at a structure to analyze such occurrences based on triggers, actions and consequences. %
Transformation is triggered by an accessibility mismatch in a representation that is necessary to achieve a work demand. In response, teams' actions consist of transformations \emph{and} an often significant amount of coordination. We characterize these into four types of patterns involving solitary ad-hoc fixes, advocacy for better standards, parallel representations, and collective assembly of information, which can be consequential for the team and the distribution of labour.

We offer a number of opportunities for the design of more accessible representational systems that are derived from our analysis. %
Our findings and perspective also reinforce the importance of team coordination, values and culture for mixed-ability teams, with regard to the sometimes mundane work of choosing representations and transforming them, for individuals or for others.

This paper contributes three main elements to the understanding of accessibility-driven information transformation in mixed-visual ability teamwork:
\begin{itemize}
    \item A structure, based on triggers, actions, and consequences, for the deconstruction of accessibility-driven transformation teamwork.
    \item A characterization of triggers, transformations, coordination, and patterns of teamwork.
    \item Identification of four areas of opportunity for the design of future collaborative representational systems.
\end{itemize}

\section{Related Work}

We use a representational lens to examine the intersection of professional knowledge work and accessibility.
We first clarify what we mean by representation and transformation, and then discuss relevant literature on accessibility in workplace settings and %
applicable research in mixed-visual teams. 

\subsection{Representation, Transformation, and Application to Work}

In this paper, we synthesize concepts from prior work~\cite{devries-LearningExternalRepresentations-2012, zhang-NatureExternalRepresentations-1997} to characterize a \emph{representation} as \emph{any structure in the environment that encodes information or data and allows people to interact with this information or data}. Such representations are fundamental tools for knowledge work~\cite{Chalmers_1999, Suchman1995, Hutchins_1995, scaifeExternalCognitionHow1996a, Clancey_1995, Jones_Chin_1998, Nielsen_2003} used by employees to manage information and carry out tasks~\cite{Suchman1995, Hutchins_1995, Erickson_Danis_Kellogg_Helander_2008}. These representations can take many forms from documents to slideshows and visualizations, and transform into other forms while having the same body of information~\cite{Ewenstein_Whyte_2009,binksRepresentationalTransformationsUsing2022,Eilam_Ofer_2018}.

The multiplicity of representational forms of information is particularly critical for accessibility in mixed-visual ability work, because workers might transform one type of representation into an accessible form (like an alternative text for screen-reading~\cite{Jung_Mehta_Kulkarni_Zhao_Kim_2022}). 
We %
argue that for mixed-visual teams, information transformation is not just an implicit feature of work~\cite{Barley_2015}, but an important mechanism through which accessibility is broken
and collectively repaired.

\subsection{Addressing Accessibility in Workplace Settings}

Currently, workplaces use accommodations as a primary means to address inclusion. For example, assistive technologies have been incorporated into workplace environments following guidelines by Dickson et al.~\cite{Dickson_Moore_Bruyere_2000} and the American Foundation for the Blind~\cite{Workplace_Technology_Study}. Even so, employers often lack knowledge about job accommodations for BLV employees~\cite{McDonnall_O'Mally_Crudden_2014}. %
Researchers 
describe this misalignment between organizational commitments and BLV workers' lived realities as the accessibility paradox~\cite{Marathe_Piper_2025} and
suggest the co-creation of work practices and the need for more accessible ecosystems~\cite{Pandey_Kameswaran_Rao_O'Modhrain_Oney_2021}.

Without proper support in place, BLV employees have to adapt existing workplace infrastructure and implement workarounds~\cite{Branham_Kane_2015}, which hinders their contribution and agency~\cite{Mörike_Kiossis_2024}.
This is particularly applicable in inaccessible software development environments, project management tools, and collaborative platforms~\cite{Cha_Figueira_Ayala_Edwards_Garcia_VanDerHoek_Branham_2024, Das_IdeationDisabledProfessionals, Huff_WorkExperiencePVI}, forcing BLV employees to use do-it-yourself (DIY) tools to simply do their work~\cite{Cha_DIYDilemma}. This form of ``working to work'' places burdens on BLV workers~\cite{PeterBloom_HackingWork, Cha_DIYDilemma}. 
To alleviate such burdens, researchers have proposed that workplaces could benefit from incorporating an interdependence framework that emphasizes collaborative accessibility, viewing access as a shared responsibility instead of an individual one~\cite{Bennett_Brady_Branham_2018, Xiao_Bandukda_Angerbauer_Lin_Bhatnagar_Sedlmair_Holloway_2024, Yildiz_Subasi_2023}. 

We contribute to this body of literature and extends the inquiry to collective, moment-to-moment practices of the entire mixed-visual teams. We examine beyond individual tools~\cite{Cha_DIYDilemma, Perera_GenAIProductivity, Branham_Kane_2015}, but also how teams collaboratively adapt and transform the everyday information representations (e.g., spreadsheets, documents, diagrams) that enable knowledge work. We argue that this ``invisible work'' is not just about tool-building, but is fundamentally about the constant, collaborative transformation of shared information.

\subsection{Information Access in Teams with Mixed-Visual Abilities}
\label{sec:RW_MixedVisualAccess}

Previous research on supporting mixed-visual abilities shows that accessible information representations are important for inclusive collaboration. In programming, tools like CodeWalk \cite{Potluri_Pandey_Begel_Barnett_Reitherman_2022}, which use sounds and speech to convey code review actions as alternative modalities to reduce coordination burden on BLV developers %
along with plugins to aid in code navigation~\cite{Albusays_CodeNav} and custom scripts for debugging~\cite{saben2024enablingblvdevelopersllmdriven}. 
In collaborative writing, such auditory representations and cues can improve collaboration awareness and coordination for BLV writers using digital tools \cite{Das_Gergle_Piper_2019, Das_Piper_Gergle_2022, Das_McHugh_Piper_Gergle_2022, Lee_Zhang_Herskovitz_Seo_Guo_2022}. 
Similarly, researchers advocated for accessible data visualization that accommodates diverse sensory modalities~\cite{Kim2021} and 
leverages tactile graphics to make visual representations tangible for BLV team members~\cite{deGreef_Moritz_Bennett_2021}. %
The need for accessible information representations extends beyond specific domains. 
Studies have identified barriers for BLV users when interacting with diagram editing tools in a collaborative setting \cite{Fan_Glazko_Follmer_2022, Das_Stangl_Findlater_2024}. 
Even routine activities like meetings can be challenging for BLV professionals due to the lack of accessibility features, often requiring them to spend extra effort to participate fully \cite{Cha_Figueira_Ayala_Edwards_Garcia_VanDerHoek_Branham_2024}. 
More recently, researchers explored how generative AI could enhance productivity applications~\cite{Perera_GenAIProductivity}.  

However, existing work often seeks to adapt existing tools to meet individual needs, overlooking the dynamic interplay of information representations within mixed-visual ability teams. 
They often frame accessibility as a one-directional translation from a visual representation to an alternative non-visual one for BLV users. 
For example, while audio descriptions benefit BLV users, they may hinder communication and shared understanding with sighted colleagues who rely on visual representations~\cite{ahlander}. 
Moreover, any new technology must support existing workflows by BLV users and allow for verification, showing that a simple translation is not a complete solution~\cite{Perera_GenAIProductivity}. 
Even when technically accessible representations are available, challenges can arise from the need to educate sighted colleagues about their limitations and importance~\cite{Wahidin_Waycott_Baker_2018}.

In our study, we aim to shift the focus from individual tools and domain-specific challenges to the core collaborative processes of how information representation is constructed, transformed, and negotiated to bridge differing visual abilities within mixed-visual teams. %
\section{Methodology}

To understand %
how mixed-visual teams %
transform information for accessibility to enable work, 
we conducted a diary study and follow-up individual interviews with 23 participants (from five teams) with diverse visual abilities to capture in-situ, moment-to-moment practices. 
We also conducted two focus group sessions involving seven participants (from two other teams). These sessions were alternative time-flexible participation options for those who could not take part in the diary study. This method enabled us to learn about people's experiences through a group discussion format~\cite{Wilkinson_1998}. This study was approved by our university's ethics board.

\subsection{Diary Study Procedure}

The diary method is a qualitative approach for capturing participants' interaction experiences over time~\cite{Lischetzke_Könen_2021}. We chose this method because it helped us capture information on how representations evolved in real-world work contexts over time, and allowed participants to reflect on their practices at their own pace. Alternatives like contextual inquiry, although they could provide deeper insights, we considered impractical for our study as they would require us, researchers, to be embedded in multiple workplaces, and would likely be considered intrusive by workplaces where there are strong expectations of privacy.

We asked participants to gather data over five working days. We chose a period of five work days so we could capture repeating patterns of transformation, and/or multiple cases of representation transformation as would be typical of a work week. Although a longer study period could have captured more repeating patterns of data transformations,  we recognized that documenting these diaries adds additional labour, especially for participants already managing accessibility barriers. As such, extending the study risked causing fatigue for our participants, and likely lower data quality. Our five-day diary study consisted of three main phases: pre-study brief, diary entries, and individual post-study interviews. 

\subsubsection{Pre-Study Brief}
We conducted a 30-minute pre-study brief session, wherein participants provided consent, and were introduced to the concept of information representation with examples. In this session, we also provided instructions on how to document their experiences and submit diary entries. %

\subsubsection{Gathering Diary Entries}
Over five working days, participants individually documented their experiences with information representations. They documented the representations they used at work, capturing a spectrum of encounters ranging from successful use to instances where representations remained inaccessible. 
They recorded contextual information related to these representations, including collaborators involved in using these representations, purpose of the representations, participant's role in using or modifying the representations, and any challenges faced or opportunities for improvement related to using representations for teamwork. 

Participants could submit their diary entries in 
their preferred format (e.g., as word-processing documents, voice memos, screenshots, text messages, or emails).  The first author checked in daily with individual participants via their preferred method (email, text message, or call/online meeting) to address questions and concerns, and to remind participants to gather data throughout the full period.
Multiple members in a team often gathered data about the same representations and contexts of use, allowing us to better understand the different perspectives on the same information representation.

\subsubsection{Post-Study Semi-structured Interviews with Individual Participants}

Following the completion of the diary study, participants took part in one-on-one post-study interviews (50-90 minutes). The aim was to gather additional reflections on their experience with information representations. 

Each interview began with the collection of demographic and background information, including age, gender, educational and professional background, and visual ability. 
Next, participants answered a set of interview questions that focused on the participants' experiences with representations and workflows based on their diary entries. 
Key topics included the factors affecting the choice of representation transformations, the impact of visual abilities on these choices, teams' interactions with the representations, and strategies with the associated accessibility challenges. %

\subsection{Additional Focus Group Procedure} %

To accommodate 
teams with tighter schedules or those unable to commit to a week-long diary study, we offered time-flexible, alternative focus group interviews~\cite{Wilkinson_1998, Baker_2023}. %
A priori, we were unsure if the focus groups would offer a different perspective than the diary study or just confirm the same findings. Section 4.2 discusses how the findings from the two methods relate. 
Participants first completed a demographic survey asynchronously. We also asked them to reflect on working with information representations and to send us examples of representations they used at work in advance of the session. They then each joined a synchronous remote focus group session (one-hour) facilitated by the first author. Participants started by elaborating on their experiences with inaccessible representations, and then participated in a group discussion about the same topics we covered in the individual interviews.

\subsection{Participants} 
We recruited 30 participants across seven teams, with 23 individuals (14 BLV, 9 sighted) from five teams participating in the diary study with individual interviews, and the remaining 7 (4 BLV, 3 sighted) from two teams joining focus groups. 
Participant labels throughout the paper include their visual abilities in the subscripts (``S'' for sighted, ``LV'' for low vision, ``LB'' for legally blind, and ``TB'' for totally blind). 
Most participants had post-secondary education; three (P5\sub{LV}, P17\sub{LB}, P30\sub{LB}) completed high school. 
In accordance with ethics, we do not reveal participants' specific professional roles to safeguard their anonymity, given the risks of employment discrimination and the small number of BLV professionals in specific positions. 
P3\sub{TB}, P7\sub{LB}, P8\sub{LB}, P15\sub{TB} are in leadership positions. %
Teams T1 and T7 are in the legal profession, T5 is in consulting, T2, T3, and T4 are non-profit, and T6 is in academia. 
P13\sub{TB}, P14\sub{LV}, and P18\sub{TB} held positions explicitly dedicated to technical accessibility compliance or remediation. 
Table \ref{tab:p_demo} summarizes participants' other demographics. 
Teams T1-5 participated in the diary study (including individual interviews) and T6-7 in the focus groups. 
All participants were also given the opportunity to review their quoted statements in the manuscript and remove any details that might compromise their anonymity.

\aptLtoX{\begin{table*}[htbp]
  \centering
  \caption{Participants demographics.} %
  \label{tab:p_demo}
    \begin{tabular}{|p{1.3em}|p{2.1em}|p{1.3em}|p{4.8em}|p{6.1em}|p{4.1em}|p{17.9em}|}
    \multicolumn{7}{c}{\textbf{(a) Diary study (with individual interviews) participants (T1-T5).}}\\
    \hline
    \textbf{TID} & \textbf{PID} & \textbf{Age} & \textbf{Gender} & \textbf{Visual Ability} & \textbf{Onset} & \textbf{Assistive Technologies} \\
    \hline
    T1    & P1\sub{TB}    & 40s   & Male  & Totally blind & Childhood & Screenreaders, Slate and stylus \\
    \hline
    T1    & P2\sub{S}    & 60s   & Female & Sighted &       &   \\
    \hline
    T2    & P3\sub{TB}    & 60s   & Female & Totally blind & 20s   & Screenreader \\
    \hline
    T2    & P4\sub{S}    & 50s   & Female & Sighted &       &   \\
    \hline
    T2    & P5\sub{LV}    & 80s   & Female & Low vision & 60s   & Magnification, Large prints \\
    \hline
    T2    & P6\sub{S}    & 20s   & Female & Sighted &       &   \\
    \hline
    T3    & P7\sub{LB}    & 30s   & Female & Legally blind & Childhood & Screenreaders, High contrast, Magnification \\
    \hline
    T3    & P8\sub{LB}    & 50s   & Female & Legally blind & Teenage & Screenreader \\
    \hline
    T3    & P9\sub{S}    & 30s   & Female & Sighted &       &   \\
    \hline
    T3    & P10\sub{S}   & 20s   & Female & Sighted &       &   \\
    \hline
    T3    & P11\sub{LB}   & 30s   & Female & Legally blind & Birth & Screenreaders, Braille \\
    \hline
    T3    & P12\sub{LB}   & 50s   & Female & Legally blind & Teenage & High contrast, Magnification, Screenreader \\
    \hline
    T3    & P13\sub{TB}   & 40s   & Male  & Totally blind & Birth & Screenreaders \\
    \hline
    T3    & P14\sub{LV}   & 40s   & Female & Low vision & 20s   & Screenreaders \\
    \thickhline
    T4    & P15\sub{TB}   & 50s   & Female & Totally blind & Birth & Screenreaders, Braille \\ 
    \hline
    T4    & P16\sub{TB}   & 40s   & Non-binary & Totally blind & Birth & Screenreaders, Braille \\
    \hline
    T4    & P17\sub{LB}   & 60s   & Female & Legally blind  & 30s   & Screenreader, Hearing aid \\
    \hline
    T4    & P18\sub{TB}   & 40s   & Female & Totally blind & Birth & Screenreaders, Slate and stylus, BrailleNote \\
    \hline
    T4    & P19\sub{S}   & 40s   & Female & Sighted &       &   \\
    \hline
    T4    & P20\sub{S}   & 40s   & Male  & Sighted &       &   \\
    \hline
    T5    & P21\sub{LB}   & 40s   & Male  & Legally blind & Childhood & Screenreaders, High contrast, Magnification \\
    \hline
    T5    & P22\sub{S}   & 20s   & Female & Sighted &       &   \\
    \hline
    T5    & P23\sub{S}   & 20s   & Male  & Sighted &       &   \\
    \hline
\multicolumn{7}{c}{\textbf{(b) Focus group participants (T6-T7).}}\\
    \hline
    \textbf{TID} & \textbf{PID} & \textbf{Age} & \textbf{Gender} & \textbf{Visual Ability} & \textbf{Onset} & \textbf{Assistive Technologies} \\
    \hline
    T6    & P24\sub{LB}   & 50s   & Female & Legally blind & 40s   & Screenreader, High contrast \\
    \hline
    T6    & P25\sub{S}   & 50s   & Female & Sighted &       &   \\
    \hline
    T6    & P26\sub{S}   & 40s   & Male  & Sighted &       &   \\
    \hline
    T6    & P27\sub{S}   & 20s   & Female & Sighted &       &   \\
    \hline
    T7    & P28\sub{LB}   & 40s   & Female & Legally blind & 30s   & Magnification, High contrast \\
    \hline
    T7    & P29\sub{LV}   & 40s   & Female & Low vision & 20s   & Magnification, High contrast \\
    \hline
    T7    & P30\sub{LB}   & 20s   & Male  & Legally blind & Childhood & Magnification, High contrast \\
    \hline
    \end{tabular}%
\end{table*}}{
\begin{table*}[htbp]
  \centering
  \caption{Participants demographics.} %
  \label{tab:p_demo}
  
  \begin{subtable}{\textwidth}
    \centering
    \small
    \caption{Diary study (with individual interviews) participants (T1-T5).}
    \label{tab:p_demo_diary}
    \begin{tabular}{|p{1.3em}|p{2.1em}|p{1.3em}|p{4.8em}|p{6.1em}|p{4.1em}|p{17.9em}|}
    \thickhline
    \textbf{TID} & \textbf{PID} & \textbf{Age} & \textbf{Gender} & \textbf{Visual Ability} & \textbf{Onset} & \textbf{Assistive Technologies} \\
    \thickhline
    T1    & P1\sub{TB}    & 40s   & Male  & Totally blind & Childhood & Screenreaders, Slate and stylus \\
    \hline
    T1    & P2\sub{S}    & 60s   & Female & Sighted &       &   \\
    \thickhline
    T2    & P3\sub{TB}    & 60s   & Female & Totally blind & 20s   & Screenreader \\
    \hline
    T2    & P4\sub{S}    & 50s   & Female & Sighted &       &   \\
    \hline
    T2    & P5\sub{LV}    & 80s   & Female & Low vision & 60s   & Magnification, Large prints \\
    \hline
    T2    & P6\sub{S}    & 20s   & Female & Sighted &       &   \\
    \thickhline
    T3    & P7\sub{LB}    & 30s   & Female & Legally blind & Childhood & Screenreaders, High contrast, Magnification \\
    \hline
    T3    & P8\sub{LB}    & 50s   & Female & Legally blind & Teenage & Screenreader \\
    \hline
    T3    & P9\sub{S}    & 30s   & Female & Sighted &       &   \\
    \hline
    T3    & P10\sub{S}   & 20s   & Female & Sighted &       &   \\
    \hline
    T3    & P11\sub{LB}   & 30s   & Female & Legally blind & Birth & Screenreaders, Braille \\
    \hline
    T3    & P12\sub{LB}   & 50s   & Female & Legally blind & Teenage & High contrast, Magnification, Screenreader \\
    \hline
    T3    & P13\sub{TB}   & 40s   & Male  & Totally blind & Birth & Screenreaders \\
    \hline
    T3    & P14\sub{LV}   & 40s   & Female & Low vision & 20s   & Screenreaders \\
    \thickhline
    T4    & P15\sub{TB}   & 50s   & Female & Totally blind & Birth & Screenreaders, Braille \\ 
    \hline
    T4    & P16\sub{TB}   & 40s   & Non-binary & Totally blind & Birth & Screenreaders, Braille \\
    \hline
    T4    & P17\sub{LB}   & 60s   & Female & Legally blind  & 30s   & Screenreader, Hearing aid \\
    \hline
    T4    & P18\sub{TB}   & 40s   & Female & Totally blind & Birth & Screenreaders, Slate and stylus, BrailleNote \\
    \hline
    T4    & P19\sub{S}   & 40s   & Female & Sighted &       &   \\
    \hline
    T4    & P20\sub{S}   & 40s   & Male  & Sighted &       &   \\
    \thickhline
    T5    & P21\sub{LB}   & 40s   & Male  & Legally blind & Childhood & Screenreaders, High contrast, Magnification \\
    \hline
    T5    & P22\sub{S}   & 20s   & Female & Sighted &       &   \\
    \hline
    T5    & P23\sub{S}   & 20s   & Male  & Sighted &       &   \\
    \thickhline
    \end{tabular}%
  \end{subtable}
  
  \vspace{1em} %
  
  \begin{subtable}{\textwidth}
    \centering
    \small
    \caption{Focus group participants (T6-T7).}
    \label{tab:p_demo_focus}
    \begin{tabular}{|p{1.3em}|p{2.1em}|p{1.3em}|p{4.8em}|p{6.1em}|p{4.1em}|p{17.9em}|}
    \thickhline
    \textbf{TID} & \textbf{PID} & \textbf{Age} & \textbf{Gender} & \textbf{Visual Ability} & \textbf{Onset} & \textbf{Assistive Technologies} \\
    \thickhline
    T6    & P24\sub{LB}   & 50s   & Female & Legally blind & 40s   & Screenreader, High contrast \\
    \hline
    T6    & P25\sub{S}   & 50s   & Female & Sighted &       &   \\
    \hline
    T6    & P26\sub{S}   & 40s   & Male  & Sighted &       &   \\
    \hline
    T6    & P27\sub{S}   & 20s   & Female & Sighted &       &   \\
    \thickhline
    T7    & P28\sub{LB}   & 40s   & Female & Legally blind & 30s   & Magnification, High contrast \\
    \hline
    T7    & P29\sub{LV}   & 40s   & Female & Low vision & 20s   & Magnification, High contrast \\
    \hline
    T7    & P30\sub{LB}   & 20s   & Male  & Legally blind & Childhood & Magnification, High contrast \\
    \thickhline
    \end{tabular}%
  \end{subtable}
\end{table*}}

We recruited participants by reaching out to local organizations, by sharing our recruitment materials via mailing lists, newsletters, and social media groups, and via snowball sampling. Interested individuals reached out to us. %
Upon checking if the individual met our inclusion criteria (adult, professional knowledge worker who engages in collaborative activities, and working in a team where at least one colleague is blind or has low vision), we contacted their workplace supervisor. This was to secure organizational consent, ensuring we had permission for participants to gather data about their collaborative work during company time. After receiving this consent, we asked the individual and supervisor to help advertise the study to other members of their team.
We recruited for the diary study first, then for focus groups. Each participant received CAD\$50 for participating in the diary study or CAD\$20 for the focus group. 

\subsection{Data Analysis}
\label{sec:dataAnalysis}

In total, we collected 404 diary entries, 209 unique 
data representation examples, 29 hours and 42 minutes of interview recordings, and two hours of focus group discussion recordings. We transcribed all recordings and analyzed them using open coding followed by deductive coding~\cite{Adams_Lunt_Cairns_2008, Muller_2014}. We grouped the coded data into themes using thematic analysis~\cite{Braun_Clarke_2023}. 

The first author began by open-coding the diaries to capture low-level, representation-oriented concepts such as the specific types of representations used, their attributes, and the immediate challenges encountered (e.g., ``merged cells,'' ``obstructed visual content''). %
Interview transcripts were then open-coded to surface higher-level social experiences and influencing factors (e.g., ``asking for help,'' ``open workplace culture''). 
To develop an initial codebook, the first author analyzed diaries and transcripts from two randomly selected teams. 
The full research team then held a series of iterative meetings to discuss these initial findings, collaboratively refining and consolidating the codes (e.g., codes about low-level challenges were grouped under higher-level codes like ``inconsistent structure'' and ``low-vision challenges''). 
Once this initial codebook was established, the first author proceeded to code the remaining diaries and transcripts, with regular team meetings to resolve any ambiguities and ensure consistency. When a new code was added, the first author revisited all diaries and transcripts to identify relevant evidence that supported it. 
The focus group transcripts were then deductively coded.

We then iterated on the analysis with particular focus on the processes through which information was transformed by team members. %
The first author revisited the data to identify cases in which team members encountered representational incompatibilities and took a series of steps to resolve them, and coded the
triggers that prompted representational changes, the actions taken by team members, and the resulting consequences for each case. 
Throughout the analysis, the research team engaged in iterative discussions about newly identified codes and restructured the codebook. %
This led to the synthesis of the sequences of triggers, actions and consequences into patterns of teams' practices. The final codebook has 96 codes. 
We continued looking for relationships across high-level codes to refine the broader themes 
(i.e., by comparing the triggers from different patterns, we realized a progression from reactive triggers to proactive which influenced how teams respond, 
leading to the development of a theme, ``From Reactive to Proactive Triggers of Transformation.'')
Through more iterations and discussion, we agreed on four overarching themes, which map to the headings in 
Section~\ref{sec:findings}. %

\section{Representations and Cases of Transformation}
\label{results:repAndCases}

We report our analysis from the bottom up. In this section we provide a quick overview of the data collected, and %
a selection of representative cases. Section~\ref{sec:findings} builds upon these cases to describe higher-level themes.

\subsection{Work Demands and Representations}
\label{results:workDemandAndRep}
The teams that we studied carried out varied work and had a wide variety of team objectives and deliverables, such as creating documents for external entities (e.g., application documents---T1-2, 4), spreadsheets for tracking financial information (T3-5), processing information from a centralized company database (T1, T3, T5), presentations (T3, T5-6), and videos for publicity (T4). Accomplishing team goals required a diverse set of representations including PDF documents, slideshows, and databases, which were sometimes also deliverables themselves. Table~\ref{tab:representations} lists the different types of representations that they used, and the number of instances of encountering them %
in the collected data. ``Instances'' denotes unique representation artifacts; if multiple members referenced the same representation in different diary entries, it was counted once. We excluded individual e-mails, short text messages, calls and meetings, which, although could be considered representations, fall below the granularity of our analysis. 

\begin{table}[h!]
\centering
\caption{Types of information representations and their usage across teams.}
\label{tab:representations}
\begin{tabular}{@{}lrl@{}}
\toprule
\textbf{Type of Representation} & \textbf{Instances} & \textbf{Teams} \\ \midrule
Documents (Word, PDF)     & 89                 & T1-7 \\
Tables or Spreadsheets          & 22                 & T1-5          \\
Slideshow Presentations         & 17                 & T3, T5-6                  \\
Databases    & 10                 & T1, T3, T5                  \\
Diagrams or Visualizations      & 9                  & T3-5, T7              \\
Podcasts or Videos              & 4                  & T3-4                      \\ \bottomrule
\end{tabular}
\end{table}

\subsection{Cases}
\label{results:casesTAC}

\newcommand{\trig}{\textbf{TR}}
\newcommand{\trigwd}{\textbf{TR\textsubscript{wd}}}
\newcommand{\triginc}{\textbf{TR\textsubscript{r.inc}}}
\newcommand{\action}{\textbf{A}}
\newcommand{\actioncoord}{\textbf{A\textsubscript{coord}}}
\newcommand{\actiontrans}{\textbf{A\textsubscript{trans}}}
\newcommand{\conseq}{\textbf{C}}

As the analysis progressed, we started to identify a common structure of the accessibility matters that appeared in the data. We find it useful to analyze the data at this level of granularity, and we refer to these episodes as \emph{cases}. Our analysis of patterns in Section~\ref{sec:findings} is built upon these. Our structure for Cases is as follows: a team or an individual needs to carry out a task or subtask (the work demand---e.g., adding numbers to a table) and a mismatch between the representation being used and the perceptual needs of a team member is identified (e.g., a blind user is sent an inaccessible PDF). We call this the trigger, which has two components: the work demand and the representation incompatibility.
A trigger then results in action. There are two parts of the action: the coordination between team members (e.g., the receiving member communicates that the document is not appropriate), and the transformation in the representation (e.g., a team member converts the PDF into a well-structured Word document). Finally, the case results in consequences, both negative (e.g., frustration, additional workload) and positive (e.g., awareness and learning) on the individual, team, and the representation itself. 
Sometimes cases have multiple of the above components.

While our data includes instances where representations remained untransformed (and thus persisted as barriers), our analysis focuses on transformation. 
We identified 36 such cases from the diary study and six additional cases from the focus groups. 
The research team selected 10 representative ones (Table~\ref{tab:case_summary}) from the total 42 through an iterative review process. We first selected cases to ensure diversity across types of representation (e.g., document, spreadsheet, video), nature of triggers (reactive, proactive), different kinds of coordination (e.g., solitary work, distributed delegation), and spanning across different teams. We then deprioritized cases which shared similar structures, unless they added unique contextual detail. 
We further selected the cases that provided the clearest narrative of events for explanation. 
%
At the end, the selected cases all came from the diary study. While the focus group data provided cases that were consistent with those in the diaries, the retrospective and conversational nature of the focus group method resulted in less granular detail compared to the in-situ diary entries. Thus, we found that the two methods provided similar insights but at different levels of resolution. 
A summary of the triggers, actions, and consequences is in Table~\ref{tab:TAC_sum} (Appendix \ref{app:sum_TAC}).




%
%

\begin{table*}[htbp]

\caption{A summary of the 10 illustrative cases, deconstructed by their trigger, action, and consequence components.}
\label{tab:case_summary} 

  \small
  \centering
    \renewcommand{\arraystretch}{1.2}

  \begin{tabular*}{\textwidth}{ | l | p{0.29448\textwidth} | p{0.29448\textwidth} | p{0.29448\textwidth} | }

\thickhline
\textbf{Case} & \textbf{Triggers} & \textbf{Actions} & \textbf{Consequences} \\
\thickhline

\case{}\label{case:failedAdvocacy} & 
\textbf{Work Demand:} Team 2 needed to sign a contract with an external vendor for event logistics. \par \textbf{Incompatibility:} Document was visually inaccessible to low-vision workers. & 
\textbf{Transformation:} P4\sub{S} was forced to verbalize the contract to P5\sub{LV}. \par \textbf{Coordination:} P4\sub{S} initially negotiated with the vendor to make it: \textit{``all spaced out''}, but they \textit{``didn't listen.''}. She \textit{``gave in''} under time pressure. & 
Increased workload and emotional cost for P4\sub{S}, who described the experience as a \textit{``challenge''}. \\
\hline

\case{}\label{case:P3Individual} & 
\textbf{Work Demand:} Prepare for a meeting by reviewing three PDF email attachments. \par \textbf{Incompatibility:} PDF layouts could not be interpreted correctly by P3\sub{TB}'s screen reader. & 
\textbf{Transformation:} P3\sub{TB} used OCR to create a text-based version for herself as a solitary action. \par No coordination. %
& 
\textit{``Very time-consuming''} (P3\sub{TB}); only able to comprehend one of three documents; created an information silo. \\
\hline

\case{}\label{case:P3Proactive} & 
Identical to previous case. & 
\textbf{Transformation:} Sighted colleague provided an accessible Word version of the original PDF. \par \textbf{Coordination:} Performed by a sighted colleague unpromptedly; an example of proactive accommodation. & 
P3\sub{TB}'s workload was minimized; she was enabled to prepare; no information silo was created. \\
\hline

\case{}\label{case:bidirectional} & 
\textbf{Work Demand:} Process a shared spreadsheet for tracking clients. \par \textbf{Incompatibility:} Spreadsheet created by P11\sub{LB} had \textit{``various fonts with no level of consistency''} with visually hidden content. & 
\textbf{Transformation:} P10\sub{S} reformatted the document to make it visually clear. \par No coordination. %
& 
P10\sub{S} absorbed the workload rather than asking P11\sub{LB} to perform a visual task, noting that to P11\sub{LB}, \textit{``formatting is not important''}; highlights the bidirectional burden of accessibility. \\
\hline

\case{}\label{case:liveDescriptions} & 
\textbf{Work Demand:} Access and understand visual diagrams necessary for work tasks. \par \textbf{Incompatibility:} Information was presented in a purely visual format, inaccessible to BLV colleagues. & 
\textbf{Transformation:} P21\sub{LB} asked P23\sub{S} to help describe the diagram. \par \textbf{Coordination:} Through dynamic, real-time verbal description, P23\sub{S} \textit{``gauging [P21\sub{LB}'s] reaction''} and adapting. & 
Description might lose accuracy or fidelity, as P23\sub{S} later pointed out that
the mental map P21\sub{LB} formed was spatially inaccurate; fatigue for sighted colleagues. \\
\hline

\case{}\label{case:spreadsheetStandard} & 
\textbf{Work Demand:} Access a company-wide event schedule. \par \textbf{Incompatibility:} Spreadsheet had \textit{``merged rows and columns [that] made it difficult to navigate smoothly''} (P11\sub{LB}). & 
\textbf{Transformation:} P11\sub{LB} unmerged the cells for herself. \par \textbf{Coordination:} P11\sub{LB} then documented the process, and advocated for it to become the new team standard. & 
Initial increased workload for P11\sub{LB}, but led to increased capacity for the whole team. \\
\hline

\case{}\label{case:parallelDatabase} & 
\textbf{Work Demand:} Synchronize and access client information from a central database. \par \textbf{Incompatibility:} Database was inaccessible (\textit{“50 different tabs”}, P2\sub{S}); extracted documents also had inaccessible visual content or structures. & 
\textbf{Transformation:} P1\sub{TB} maintained a parallel, personal Word document for his daily work. \par \textbf{Coordination:} P2\sub{S} verbalized the initial data for P1\sub{TB}, and was delegated the ongoing task of transcribing his information updates back to the database. & 
Increased efficiency for P1\sub{TB}, but created a persistent silo, a fragile workflow dependent on P2\sub{S}, and extra reconciliation workload for P2\sub{S}. \\
\hline

\case{}\label{case:parallelDataModel} & 
\textbf{Work Demand:} Review and verify a data decision model. \par \textbf{Incompatibility:} Spreadsheet formatting standards and formula checking were inaccessible to P21\sub{LB}. & 
\textbf{Transformation:} P21\sub{LB} created a parallel representation of the model in Python to test and confirm the results. \par \textbf{Coordination:} P21\sub{LB} delegated the creation of the primary spreadsheet model to P23\sub{S} to be used by their team. The two discussed the model throughout. & 
Successful validation of the model, resulting in additional confidence for P23\sub{S} in the results. \\
\hline

\case{}\label{case:podcast} & 
\textbf{Work Demand:} Create a podcast video for publicity. \par \textbf{Incompatibility:} Sound editing was \textit{``very visual''} and inaccessible (P15\sub{TB}); required visual artwork and alt-text that P15\sub{TB} could not create. & 
\textbf{Coordination:} P15\sub{TB} delegated the audio editing and visual artwork tasks to a sighted editor and P20\sub{S}. \par
\textbf{Transformation:} P15\sub{TB} assembled the final video from the video components transformed by others. & 
Achieved the work outcome, but created a fragile workflow and was a \textit{``real stumbling block''} to her independence (P15\sub{TB}). \\
\hline

\case{}\label{case:slideshow} & 
\textbf{Work Demand:} Prepare an inclusive presentation for an external audience. \par \textbf{Incompatibility:} A standard slideshow had multiple incompatibilities (e.g., low contrast for P14\sub{LV}, visual navigation for P13\sub{LB}) for different members. & 
\textbf{Transformation:} Team members added multiple layers: a high-contrast theme, alt-texts (P10\sub{S}), and \textit{``page-shifting audio''} cues for slide transitions (P14\sub{LV}). \par \textbf{Coordination:} Different members contributed specialized components based on their abilities in a coordinated assembly. & 
Enabling for all presenters, but required increased coordination efforts and a dependency on a sighted colleague for some tasks. \\

\thickhline

  \end{tabular*}
\end{table*}

\section{Findings}\label{sec:findings}

From the deconstruction of all cases into trigger, action and consequence components describe in the previous section, we built a bottom-up analysis to answer our research question. Each section below describes an aspect of our answer. 
The notation ``$n$'' refers to the number of unique participants (out of the total 30).

\subsection{Transformations Simplify or Enhance Representations}
\label{finding:simplifyOrEnhance}

Transformation followed one of two mechanisms: enhancement and simplification. 
Enhancement (\textit{n}=13) is the practice of augmenting a representation to make it more accessible without sacrificing its core functionalities. E.g., in Case \ref{case:slideshow}, the team did not abandon PowerPoint slideshow; they enhanced it by layering multiple accessibility features, such as alt-text, audio cues, and a high-contrast theme. 
This aims to preserve the many functions of the original representation while extending its accessibility. 

However, accessibility-driven enhancement was often technically challenging because tools did not adequately support accessibility features.  
This was a common pain point for participants (\textit{n}=15; some quit enhancement attempts in favour of simplification due to technical barriers). %
P8\sub{LB} described their financial statements as \textit{``kind of accessible, but not really [...] There're components that I can navigate with JAWS [screen-reader], but then it stops. It doesn't go all the way [to enable access].''} %
Another example was creating accessible PDFs (P4\sub{S}), which involved specific technical skills that may not be part of standard workflows. 
P12\sub{LB} highlighted this difficulty in semantic formatting: \textit{``If you do the underlining and the bold first and then you do the heading [formatting], it erases the other formatting.''} 

Conversely, simplification (\textit{n}=19) reduces a representation to a more basic, universally robust format, often at the cost of its advanced features. 
In Case~\ref{case:parallelDatabase}, P1\sub{TB}'s transformation of a complex database into a simple, linear Word document is an example of simplification. This made the information accessible to him but stripped away the database's dynamic querying capabilities. 
As P12\sub{LB} argued, a simple \textit{``Word document with a list''} may be the most effective solution even though \textit{``it's very simple, very low tech, doesn't look great probably,''} because \textit{``it may be the thing that people buy into the most.''} %

Simplification is often a practical fallback when enhancement is not feasible. 
For example, the common act of copying information from inaccessible representations into a plain text document (shared by 11 participants) loses information fidelity (\textit{n}=9) but enables work. As P2\sub{S} stated: \textit{``It doesn't copy the same as the original [...] It's never identical.''}
This reflects the tension that teams are often forced to trade representational functionalities for accessibility.

\subsection{Patterns of Transformations, Coordination, and Their Consequences}
\label{finding:patterns}

\begin{figure*}[h!]
    \centering
    %
    \begin{tabular}{ m{0.75\textwidth} m{0.20\textwidth} }
        \centering
        \begin{subfigure}[t]{0.36\textwidth}
            \centering
            \includegraphics[width=0.86\linewidth]{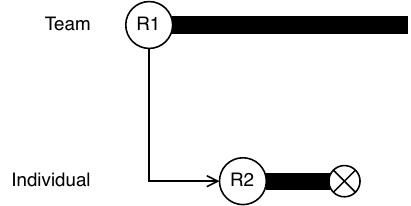}
            \caption{Pattern 1: Disposable fixes. Coordination of assistance might occur.}
            \label{fig:p1}
        \end{subfigure}\hfill
        \begin{subfigure}[t]{0.36\textwidth}
            \centering
            \includegraphics[width=0.86\linewidth]{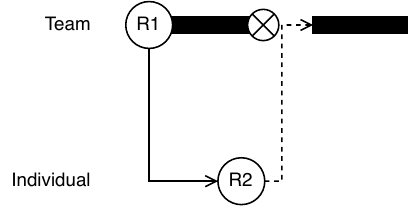}
            \caption{Pattern 2: Transformed becomes the standard. Coordination refers to advocacy.}
            \label{fig:p2}
        \end{subfigure}

        \vspace{0.8em}

        \begin{subfigure}[t]{0.36\textwidth}
            \centering
            \includegraphics[width=0.86\linewidth]{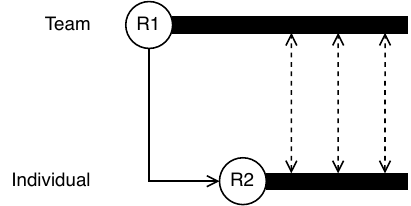}
            \caption{Pattern 3: Parallel representations. Coordination refers to synchronizing.}
            \label{fig:p3}
        \end{subfigure}\hfill
        \begin{subfigure}[t]{0.36\textwidth}
            \centering
            \includegraphics[width=0.86\linewidth]{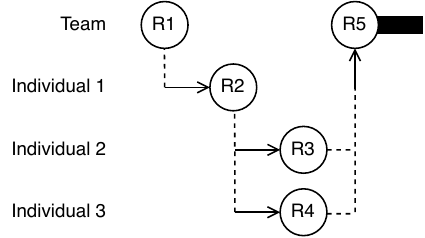}
            \caption{Pattern 4: Assembly. Coordination refers to delegating and synchronizing.}
            \label{fig:p4}
        \end{subfigure}
    \hfill
    &
    %
        \centering
        \footnotesize %
        \textbf{Legend} \\[0.8em]
        \begin{tabular}{@{}l l@{}}
            \raisebox{-0.4ex}{\Large$\circ$} & Representation \\[0.5em]
            \raisebox{0.2ex}{\rule[-0.1em]{1.2em}{0.4em}} & Work \\[0.5em]

\scalebox{1.1}{$\bm{\rightarrow}$} & Transformation \\[0.5em]

\scalebox{1.1}{$\bm{\dashrightarrow}$}       & Coordination \\[0.5em]
            $\otimes$ & Discarded \\
        \end{tabular}

    \end{tabular}

    \caption{The four patterns of transformation and coordination. Time flows from left to right.}
    \label{fig:ps}
\Description{A four-panel figure illustrating different patterns of transformation and coordination. A legend explains the symbols: a large circle is a representation, a horizontal black rectangle is work, a solid arrow is transformation, a dotted arrow is coordination, and a circle with an X is a discarded representation.

a) Pattern 1: A team uses representation R1 for its work. R1 is transformed into R2 for an individual's use. After the individual's work is done, R2 is discarded, and R1 remains the official version.
b) Pattern 2: Similar to 1, a team uses representation R1. Through advocacy, R1 is transformed into a new, accessible representation R2. The original R1 is then discarded, and the entire team adopts R2 for their work.
c) Pattern 3: A team uses a primary representation, R1. An individual transforms it into a parallel representation, R2, for their own work. Information from R2 is synchronized back to R1 continuously, and both representations are maintained.
d) Pattern 4: A team uses R1, but delegates it to multiple members to add accessibility features. R1 is given to Individual 1 to transform to R2. Then, Individual 1 delegates R2 to both Individual 2 and 3 who transform R2 simultaneously into R3 and R4. Individual 1 then synchronizes with Individual 2 and 3, and combines R3 and R4 into R5 finally. }
\end{figure*}

The detailed analysis of reported representations and transformations offers a view of how teams coordinate the labour of transformation and allows us to classify instances into four distinct, recurring patterns, illustrated in Figure~\ref{fig:ps}.

%
\subsubsection{Pattern 1: Disposable Fixes}
\label{finding:pattern1:disposableFixes}

This was the most frequently observed pattern in our data (19 cases; Case \ref{case:failedAdvocacy}, \ref{case:P3Individual}, \ref{case:bidirectional}, \ref{case:liveDescriptions} from Table~\ref{tab:case_summary}; Figure~\ref{fig:p1}), where an individual encountered an incompatible representation and transformed it to create a temporary, alternative representation for themselves 
or another colleague 
to complete a task. This can concentrate the increased workload on one individual (\textit{n}=19), as 
P8\sub{LB} expressed: \textit{``I will have to take a lot more time [... and] have to work much harder''} when dealing with inaccessible formats. 
Sighted colleagues also performed solitary actions in response to incompatible representations by their BLV teammates (\textit{n}=5). In Case \ref{case:bidirectional}, P10\sub{S} encountered a spreadsheet with inconsistent formatting from P11\sub{LB} where some text could not be visually accessed, and fixed it herself because P11\sub{LB} did not do formatting. 
Alternatively, an individual might ask for assistance with their transformation (\textit{n}=15) as a form of coordination, for example, Case \ref{case:liveDescriptions}. %
With or without coordination, the fix stayed isolated, creating an information silo (\textit{n}=10) because the transformed representation was discarded and not shared back with the team for any long-term benefit, and the original inaccessible representation remained. For example, the text converted from OCR in Case \ref{case:P3Individual} was for P3\sub{TB}'s use only. 
\subsubsection{Pattern 2: Transformed Becomes the Standard}
\label{finding:pattern2:transformedStandard}

In this pattern, a transformation evolved into a durable team standard
(nine cases; Case \ref{case:P3Proactive}, \ref{case:spreadsheetStandard} from Table~\ref{tab:case_summary}; Figure~\ref{fig:p2}). 
As with Pattern 1, one individual performed a fix, but this was followed by advocacy efforts (\textit{n}=9) to coordinate the team to align their practices, adopt the transformed representation, and discard the original, as exemplified in Case \ref{case:spreadsheetStandard}. 
This was often driven by a desire to prevent recurring problems and share a more universally accessible representation (\textit{n}=14). P13\sub{TB} articulated this motivation when he called for \textit{``a guideline if you're going to be sharing a document to everyone who's going to be accessing it.''}
This resulted in an increased capacity for the team (\textit{n}=7) in the longer term
as P6\sub{S} stated: \textit{``It's accessible to everyone: whatever level of vision that they have [...] we can all be on the same page.''}
\subsubsection{Pattern 3: Parallel Representations}
\label{finding:pattern3:parallelReps}

This pattern involves a team intentionally creating and maintaining parallel representations that co-exist with the original representation for ongoing work (eight cases; Case \ref{case:parallelDatabase}, \ref{case:parallelDataModel} from Table~\ref{tab:case_summary}; Figure~\ref{fig:p3}). This typically happened due to a persistent incompatibility from a core representation that was too central to be replaced (\textit{n}=6). 
This enabled a sustained representation that is accessible to BLV employees and free of constant patch fixes. However, it imposed more workload onto sighted colleagues who had to take on the ongoing coordination burden of synchronizing information updates (\textit{n}=9) across the parallel representations 
(e.g., P2\sub{S} in Case \ref{case:parallelDatabase} and P23\sub{S} in Case \ref{case:parallelDataModel}). 
P1\sub{TB} described this as \textit{``redundant''} but necessary.

\subsubsection{Pattern 4: Assembly}
\label{finding:pattern4:assembly}

This pattern describes how a team assembles an accessible (and sometimes multi-modal) representation through a constructive process (six cases; Case \ref{case:podcast}, \ref{case:slideshow} from Table~\ref{tab:case_summary}; Figure~\ref{fig:p4}), where multiple members transformed different parts of a shared representation and then coordinated their contributions into a final unified representation.
Case~\ref{case:slideshow} exemplifies this with a multi-stage process where members transformed a PowerPoint slideshow for their different abilities (this is represented by the sequential flow from R1, to R2 to R3 in Figure~\ref{fig:p4}). 
Some transformations could happen simultaneously, as in Case~\ref{case:podcast}, but the transformed parts required final assembly (this is represented by the flow from R2 to R3 and R4 in parallel in Figure~\ref{fig:p4}). 
This pattern enabled teamwork using shared representations, but increased coordination effort (\textit{n}=5) spent on distributing the transformation work and synchronizing how these different parts and roles fit together.

\begin{table}[t]
  \centering
  \caption{Distribution of the 42 transformation cases across the four patterns, with representative examples from Table~\ref{tab:case_summary}.}
  \label{tab:pattern_distribution}
  \begin{tabular}{l r l}
    \toprule
    \textbf{Patterns} & \textbf{\#Cases} & \textbf{From Table~\ref{tab:case_summary}} \\
    \midrule
    1. Disposable Fixes & 19 & Cases 1, 2, 4, 5 \\
    2. Transformed Becomes Standard & 9 & Cases 3, 6 \\
    3. Parallel Representations & 8 & Cases 7, 8 \\
    4. Assembly & 6 & Cases 9, 10 \\
    \bottomrule
  \end{tabular}
\end{table}

\subsection{Coordination is Negotiated Differently Across Patterns}
\label{finding:negotatingCoordination}

The coordination of transformation labour is a negotiated process; we observed how teams decide how to distribute the labour. %
Table~\ref{tab:pattern_distribution} summarizes how the cases are distributed across the four patterns. 

In Pattern~1 instances, as described in Section~\ref{finding:pattern1:disposableFixes}, there was often no coordination at all because employees transform representations themselves. However, when there was a request for assistance, coordination occured and 
the negotiation was often explicit and transactional (\textit{n}=21). %
P12\sub{LB} described: \textit{``[I] ask them to verbally tell me where [a visual element] is''} on a visual representation. 
However, this transactional nature often resulted in social cost (\textit{n}=13) %
of feeling like a \textit{``bother''} (P24\sub{LB}), experiencing a loss of autonomy, or professional cost (\textit{n}=8). 
P21\sub{LB} worried that \textit{``I can be outpaced by some sighted colleagues, even though we have very similar skills and abilities,”}
and P7\sub{LB} noted that needing constant assistance can \textit{``take away from the effectiveness of me being able to portray myself in that leadership role.''}
We infer that this is why Pattern 1 defaulted to a solitary, uncoordinated act to avoid the perceived costs of this negotiation: 
\textit{``I don't reach out and ask people for help because I didn't want them to know how bad I am''} (P17\sub{LB}).

Pattern~2 evolves from individual request to negotiation for collective change through advocacy (Section~\ref{finding:pattern2:transformedStandard}). P13\sub{TB}'s call for \textit{``a guideline if you're going to be sharing a document to everyone''} is not just a request for a fix, but a negotiation to shift the team's shared practices.

Pattern~3 is a negotiated agreement to manage a persistent barrier, but it highlights the tension between independence and support (\textit{n}=10). Although it gave BLV employees autonomy with their parallel accessible representation, it relied on a continuous and negotiated dependency (which can result in a fragile workflow; \textit{n}=9). As in Team 1, P1\sub{TB}'s work was contingent on the negotiated role of P2\sub{S} for constant information synchronization. Sighted colleagues also navigate this tension. P4\sub{S} described the tact required to avoid over-helping a BLV colleague who values her independence: \textit{``There's a fine balance between doing what is needed [...] and just going ahead and doing everything [...] I could just do it, and then she's like: `Why don't you let me?' ''}

In Pattern 4, the negotiation becomes more implicit and relational (\textit{n}=10) as teams develop awareness of each other's abilities (\textit{n}=14). 
As P23\sub{LB} said of the co-creation process: \textit{``I'm clicking the mouse and [P21\sub{LB}]'s using the keyboard. So there's [...] coordination there.''} 

Overall, in a supportive open work culture (\textit{n}=21), negotiation becomes an opportunity for building trust and understanding. P11\sub{LB} considered it as essential for \textit{``growing as a team and just being open and accommodating.''} The goal is not total independence, which can lead to isolated transformation (Pattern 1), but interdependence. As P16\sub{TB} argued, the pinnacle of success is when a team can \textit{``be interdependent and rely on each other,''} because this is what \textit{``builds resilience just by virtue of being connected.''}

\subsection{From Reactive to Proactive Triggers of Transformation}
\label{finding:reactiveOrReactive}

We did not find evidence showing the types of triggers correlate to the patterns, but
found that transformations were triggered either reactively as they arise or proactively through foresight, influencing when the trigger of a transformation is recognized and acted upon.

In a reactive coping mode, a trigger was unexpected (\textit{n}=16): the representational incompatibility %
is only discovered at the moment of use (e.g., ad-hoc transformations in Pattern 1, which often place their burdens on one individual, Section 5.2.1). 
Moreover, BLV individuals also bore the burden to inform and educate their colleagues about accessibility (\textit{n}=11). P3\sub{TB} explained when requesting an accessible format from an external provider: \textit{``PowerPoint is similar to PDF; Word works best.''} %
Pattern 2 also typically began reactively, but the act of advocacy for a more accessible standard aimed to transition the team's future action from reactive to proactive, as described in Section 5.2.2.

In contrast, when a team had proactive awareness of their colleagues' needs and how an incompatibility might unfold for them, the trigger became anticipated (\textit{n}=9), allowing the team to engage in more thoughtful and planned actions. 
In Case \ref{case:P3Proactive}, a sighted colleague demonstrated this by proactively sending P3\sub{TB} an accessible Word document alongside a PDF, eliminating P3\sub{TB}'s burden. The planned decision in Pattern 3 to support two parallel representations was also a proactive strategy to manage a persistent, known incompatibility.
Similarly, Pattern 4 (e.g., Case \ref{case:slideshow}) was motivated by the team's proactive desire to create an inclusive experience. 
P9\sub{S} described her own shift from reaction to anticipation over time: \textit{``Whenever I have to send a document in the future for anyone, it's always going to be at a certain font size [with high contrast,]''} as she gained more understanding of others' abilities. 
P23\sub{S} summarized this mature state of awareness: \textit{``You kind of know what they're capable of [...] how they prefer things.''} 

\section{Discussion}
In this paper we consider the role that representations and their transformations play in accessibility for mixed-visual ability work teams. The research uncovered fundamental aspects of the structure of work as they relate to the use of representations (e.g., documents), which is relevant for the design of new technologies that are accessible and integrate well within group work.

We first describe occurrences where a worker (sighted or BLV) needs to access information for work (Section~\ref{results:casesTAC}). 
Someone notices that the form of the information makes it difficult or impossible to access to somebody in the team (usually themselves). The combination of a work requirement and lack of representation accessibility triggers an action. 
We identified triggers when the work has no other way to continue (\emph{reactive}) and also triggers when a lack of accessibility is anticipated and acted upon (\emph{proactive}), usually for others (Section~\ref{finding:reactiveOrReactive}). 
Moreover, due to our focus on representation and the granularity of our study, the main type of action carried out in response to a trigger is a representational transformation. We have learned that these transformations sometimes \emph{enhance} the representation with features that make them more universally accessible (Section~\ref{finding:simplifyOrEnhance}). However, the transformation is more often a \emph{simplification} of the original representation, which usually produces an ephemeral alternative representation that is less likely to support multiple members of the team.

Importantly, the additional labour of transformation is accompanied by additional labour of coordination, and common representations must be agreed upon as part of how the team coordinates to complete tasks.
This is negotiated in different ways (Section~\ref{finding:negotatingCoordination}). 
The negotiation can be explicit but creates social and professional barriers that drive individuals toward solitary work; or, it evolves to become implicit toward collective change built on mutual awareness. 
We categorize the way in which transformation and coordination unfold into four patterns: disposable fixes, transformed becomes the standard, parallel representations, and assembly (Section~\ref{finding:patterns}). Each of these patterns has different consequences and reflects the team's ability to coordinate which, in turn, can result in labour unfairly piled upon BLV team members.

Our findings build directly on the literature of Branham and Kane, Cha et al., and Pandey et al. about invisible labour that BLV people perform to make inaccessible technologies usable~\cite{Branham_Kane_2015, Cha_DIYDilemma, Cha_Figueira_Ayala_Edwards_Garcia_VanDerHoek_Branham_2024, Pandey_Kameswaran_Rao_O'Modhrain_Oney_2021}. While these refer primarily to the individual burden of remediation, our analysis extends this into collaborative and representational contexts by showing how access labour is not only personal but also socially negotiated, distributed, and patterned within teams. 

\subsection{Are Some Practices Better?} 
Because our study is non-interventional, it is difficult to establish causal connections between the practices and patterns that we examined and ``better'' outcomes. Like any functional professional team, the teams that we found do their best to accomplish their goals within their means and circumstances. Additionally, team work is an ever-evolving process where teams and individuals continuously learn and adapt to be more efficient~\cite{Metz_Jensen_Farley_Boaz_Bartley_Villodas_2022, Six_2007}. 

Nevertheless, we identified circumstances and practices that participants reported as problematic, as well as trade-offs between different ways to address representational challenges. 
Our patterns can be considered as structured forms of the ``workarounds'' that Mörike and Kiossis describe as ``practices that deviate from an official pathway to a target'' for BLV workers~\cite{Mörike_Kiossis_2024}. 
The following subsections discuss the main ones and form the base for the design opportunities in Section~\ref{discussion:designOpp}.

\subsubsection{Proactive vs.\ Reactive} \emph{A priori}, teams which operate proactively will likely waste less time coordinating, and their members will be less frustrated. This is supported by observations where reactive triggers often led to the solitary actions of \emph{disposable fixes} (Section~\ref{finding:pattern1:disposableFixes}). 
The transformations that are requested and have to be coordinated in reactive scenarios might often disappear or result in minimal added labour if the person originating the representation is aware of the representational needs of the other team members (e.g., small changes such as not collapsing cells in Excel, or creating proper headings in MS Word rather than applying typographical features directly). However, reactive scenarios are likely how the awareness starts in the first place; the \emph{transformed becomes the standard} pattern advocates for moving reactive transformation to proactive (Sections~\ref{finding:pattern2:transformedStandard},~\ref{finding:reactiveOrReactive}). It is therefore not a matter of preventing teams from reacting altogether, but instead making sure that reactive triggers are identified as learning opportunities that are incorporated in the team's workflow. %

\subsubsection{Simplification vs.\ Enhancement}
More sophisticated representations typically incorporate functionality that allows workers to do more. For example, an Excel spreadsheet enables searches and filters that will allow workers to find and organize information faster than a simple list on an e-mail. However, sophisticated representations might also result in a lack of representational accessibility and cause the extra work of representational transformation. Simplifying transformations can be symptoms of complex representations with poor accessibility or of lack of training. From what we have found (Section~\ref{finding:simplifyOrEnhance}), enhancement transformations are only supported by the best or mature tools, and can be very useful because they enable the whole team to work on the same document, adapting its representation for their own and each others' abilities and preferences.

\subsubsection{Merits of the different patterns}
The coordination and representational transformation patterns that we examined are the ways in which teams naturally adapt to workflow requirements and circumstances of work (e.g., adapting to the external impositions on the format of the deliverables or to the unavailability of better representations that would allow them to work together). Hence, all patterns are necessary in different circumstances, and therefore none is inherently superior or inferior to another. Nevertheless, we suspect that teams where \emph{disposable fixes} is dominant result in much avoidable frustration and unfair labour for BLV team members,  
and are therefore something to look for and avoid. Self-advocacy and the possibility to appear less capable or professional to others plays against the visibility of this situation. Other than that, \emph{transformed becomes the standard} is likely efficient in the long term, but might often not be possible if appropriate software or training is not available; parallel representations will enable work (at the cost of having to keep different representations manually synchronized). %

\subsection{Opportunities for Design}
\label{discussion:designOpp}

Much research in HCI focuses on the design of novel and more effective accessible transformation, such as tactile alternatives for visualizations (e.g., \cite{automating_tactile, VizTouch}), new ways to access tables and diagrams for BLV users (e.g., \cite{Wang_TabularDataBLV, Zhao_Nacenta_Sukhai_Somanath_2024}), and improved alternative text for images (e.g., \cite{mack_kane_AIprompt, Jung_Mehta_Kulkarni_Zhao_Kim_2022}). From the perspective of our study, these contribute to enable enhancement transformations as ways to step up the ladder of accessibility~\cite{Zhao_Nacenta_Sukhai_Somanath_2024}. However, our findings offer additional broader context for this work. We show that the availability of a technically accessible representation does not guarantee its use. The social and logistical realities of collaborative work (triggers, actions, and consequences at a given moment) can force a team to choose simplification over enhancement, falling back to a less sophisticated but more pragmatic solution (Section~\ref{finding:simplifyOrEnhance}). 
Below we point to opportunities for the design of collaborative representational tools that support mixed-visual work teams. 
We treat them as separate, although we acknowledge they are interrelated and may overlap when applied during design.

\subsubsection{Enabling Anticipation}\label{sec:opportunities:anticipation}
Our data suggests that proactive teams anticipate representational needs, which results in less labour in transformation, less coordination and fairer distribution of labour across the team (Sections~\ref{finding:negotatingCoordination},~\ref{finding:reactiveOrReactive}). Although some existing tools are already somewhat anticipative (e.g., Word's real-time accessibility checker, PDF accessibility verifier), 
they do little to help collaborators foresee specific incompatibilities that will matter for their teammates. 
Future systems could extend these by incorporating team accessibility profiles that capture members' preferences or needs. For example, if one member typically works in Word and another avoids tightly spaced PDFs, a system preparing an outgoing document could inform the author that the current format will likely hinder colleagues and offer to convert it before circulation. 
This builds on prior work on social translucence~\cite{Erickson_Kellogg_SocialTranslucence} and interdependence in accessibility~\cite{Bennett_Brady_Branham_2018}, and makes representational consequences known early to move from reactive error-checking to proactive collaboration. 


\subsubsection{Supporting Flexible, Multi-modal Representations}\label{sec:opportunities:multimodal}
Transformations are not necessary if the representations are adaptive and sufficiently flexible, saving time and coordination effort. In our experience, this can happen in two main ways. On one side, software can use a sophisticated overlay of functionalities that support access to the underlying representation (i.e., from the \emph{Assembly} pattern---Section~\ref{finding:pattern4:assembly}---we learned about cases where a team co-constructed multi-modal representation through coordinated efforts like Case \ref{case:slideshow}). %
Prior work already explores domain-specific multi-modal access, such as auditory programming tools~\cite{Potluri_Pandey_Begel_Barnett_Reitherman_2022, Albusays_CodeNav} 
and touch-and-audio-based diagrams~\cite{Zhao_Nacenta_Sukhai_Somanath_2024}, and these efforts could serve as starting points for more general-purpose co-editing tools. 
This approach is typical of mature software and is likely to grow with the increased interpretive capability of generative artificial intelligence (GenAI) technologies (e.g.,~\cite{Glazko_Yamagami_Desai_Mack_Potluri_Xu_Mankoff_2023, Perera_GenAIProductivity}, BeMyAI).

Alternatively, careful separation of information and its format or modality can also be powerful (e.g., the underlying data is separate from its visual specification using Vega-Lite~\cite{2017-vega-lite}). 
This helps address the cause of simplification transformations (Section~\ref{finding:simplifyOrEnhance}). For example, in cases like Case \ref{case:parallelDatabase}, where a database was reduced to a linear Word document, the team was manually removing an inaccessible representation layer to get to the core semantic information. 
An \emph{amodal representation}, as advocated by Lewis~\cite{lewis2017representation}, is exemplified by the early days of HTML, when most tags in a web page were used to denote the semantic structure, not the intended visual rendering. 
Similarly, a spreadsheet or document would remain a single shared artifact, but each team member could render it in the modality or format that best supports their workflow. 

\subsubsection{Supporting and Guiding Collaborative Patterns}
While the previous opportunities focus on anticipating issues or adaptive representations, here we consider how systems can detect and guide the patterns teams fall into during transformation, 
which have different consequences (Sections~\ref{finding:patterns},~\ref{finding:negotatingCoordination}).
The current popular tools to manage, support and record workflows such as Monday.com or MS Teams, or extensions of current office work suites could potentially detect undesirable patterns and encourage workflows that better support mixed-ability work.
This could be facilitated through an intelligent assistant (e.g., Zoom AI Companion, Microsoft Copilot) that monitors collaborative activities in real-time. 
For example, if the assistant detects several members independently reformat the same spreadsheet to improve screen-reader navigation (\textit{disposable fixes} pattern), it could intervene to suggest adopting the most accessible version as a shared template (transiting to \textit{transformed becomes the standard} pattern). 

\subsubsection{Supporting Negotiation of Access and Trust Building}
Although it might be tempting to try to minimize transformation and coordinated work at all costs (as suggested by the opportunities in Subsections~\ref{sec:opportunities:multimodal} and~\ref{sec:opportunities:anticipation}), P11\sub{LB}, P16\sub{TB} and P25\sub{S} explicitly noted that the coordination necessary for transformation is also an opportunity to build trust. Systems will have to find the balance between eliminating tedious and frustrating work while simultaneously enabling social opportunities to build trust and understand each other's abilities. This echoes earlier recommendations to reduce frictions without eliminating relational benefits~\cite{Frydlinger_Vitasek_Bergman_Cummins_2021}. 
One possibility is to integrate low-barrier help-seeking within representations themselves. For example, right-clicking an image or complex table could present an option to ``ask a colleague for help with this,'' automatically attaching the problematic element and prompting the colleague with brief, research-informed accessibility guidance. Similar ideas appear in work on accessibility snippets for programming~\cite{Pandey_snippetsTooltipsUIDocs}, which could serve as starting points for general co-editing tools. The resulting annotated transformation (e.g., alt-text description, data overview) could remain attached to the representation, reducing the need for recurrent negotiations.

\subsection{Opportunities for Teams and Organizations}
The previous subsection points to opportunities for the design of representations and technology to improve how mixed-ability teams work together. However, it would be naive to assume that the main issue is technological. The diaries from all teams contain examples that show that the ultimate adaptations are social and cultural. Teams cannot function well without awareness of each other or a workplace \textit{``culture of inclusion''} (P7\sub{LB}) that recognizes values of fairness and mutual support. 

Our findings highlight that the representational transformations and their coordination are a form of the \emph{invisible work} of accessibility, a concept well-established by scholars in the domain~\cite{Mörike_Kiossis_2024, Branham_Kane_2015, Cha_DIYDilemma}. This labour is a specific type of awareness and coordination work, as defined in the computer-supported cooperative work (CSCW) field~\cite{AwarenessCoordinationInWork,Gutwin_Greenberg_2000}, that is required to maintain a shared understanding in the face of representational breakdowns. The most effective teams are those that have developed what Mia Mingus terms \emph{access intimacy}: a deep, often unspoken, understanding of each other's access needs~\cite{Mingus_2011}.

Notwithstanding how pervasive social solutions to the problem of inaccessible representations were, our participants shared practices and culture changes that can significantly improve teamwork in mixed-ability teams. We would like to highlight the two most obvious ones. First, teams can create regular, low-stakes opportunities to discuss accessibility by being \textit{``open and accommodating so that others can approach you” }(P11\sub{LB}). Simple practices, such as beginning projects with team discussions about preferred formats and tools, can normalize inquiring about one's access needs, and reduce the social cost for individuals who would otherwise need to ask for help repeatedly. Similarly, knowledge-sharing practices, like the JAWS user support group in Team 3, proved effective in building collective capability. This is consistent with previous work (e.g.,~\cite{Cha_DIYDilemma, Momotaz_screenReader}) but seems particularly relevant for the issue of representation.

Second, organizations should recognize that the burden of transformation should not fall on individual employees, whether BLV or sighted. The prevalence of Pattern 1 signals systemic failure, an exacerbated symptom of \emph{invisible work}~\cite{Branham_Kane_2015}. 
When teams repeatedly encounter inaccessible representations from external sources, as in Case \ref{case:failedAdvocacy}, organizations should support them in advocating for change with external parties. 
P16\sub{TB}'s experience illustrates that such advocacy can make a \textit{“company [...] a lot more aware.”} 
By treating accessibility as a collective, strategic priority, organizations can shift the burden from individual coping to a shared responsibility for creating a more inclusive and effective workplace.

\subsection{Study Limitations and Future Work}

Although we engaged groups from different professional backgrounds, it remains difficult to assess how representative our sample is of the broader composition of mixed-visual ability teams across different industries or geographic locations. %
We acknowledge that the non-profit nature of the three teams may have fostered greater accessibility awareness, even though only three participants held explicit accessibility roles. Despite this, we still observed representational breakdowns in these teams from the same triggers as others, indicating transformation remains crucial for mixed-visual ability teamwork. 
It is also possible that our cases and patterns of teams' practices are not exhaustive. %
For example, factors such as different organizational cultures, power relationships or lack of training might contextualize triggers of transformation and give rise to other patterns. 
In our analysis, we considered a broader context for each case but we do not report such details because such factors are well-known in literature (e.g., ~\cite{Wu_DataEverywhere, Wong_Hurbean_Davison_Ou_Muntean_2022, Karunakaran_Reddy_2012, Suchman1995, Reddy2006}) or we judged them less relevant. 
Future studies can help extend or even validate these transformation practices in different professional contexts to build a more comprehensive understanding of information accessibility problems as a research community. 

Additionally, there is a natural bias towards identifying problems in our findings as participants more readily discussed barriers to accessibility than successes like smooth collaboration or effective accessible representations. This means that the positive aspects of current practices may not be as thoroughly represented in our results, affecting the balance of the study's findings. Future work can explore co-design of representations with mixed-visual teams to better focus on positive aspects with an ability-based lens~\cite{Wobbrock_Kane_Gajos_Harada_Froehlich_2011}.

Moreover, we conducted this study during a period of transition in workplace tools (July 2024) with the growing influence of GenAI (e.g., BeMyAI, JAWS Picture Smart). 
Although these tools were already available, our participants commented that organizational policies around work information often prevented their use (also mentioned in~\cite{Perera_GenAIProductivity}). 
Despite the evolving GenAI landscape, we believe our findings will be helpful for shaping the future use of these tools. It is important to build on these insights to ensure that accessibility remains a priority as new technologies transform work environments. We also need to conduct further studies to understand the impact of such transformations on accessibility and collaboration. Future work should explore how these technologies can enhance real-time adaptability in mixed-visual teams and ensure that GenAI-powered representations can be both personalized and aligned with team goals.

Finally, our analysis examines processes and functional consequences of transformations, not their quality of outcomes or participants' subjective attitudes (e.g., satisfaction). We deliberately did not treat patterns as inherently positive or negative because such judgments can obscure the contextual reasons for people's behaviours (e.g.,~\cite{Smyth_Kumar_Medhi_Toyama_2010}). 
Our data show trade-offs: i.e., a transformation may be technically inefficient (high workload) but socially beneficial (building trust). Future work should investigate these to determine how (perceived) outcome quality affects long-term collaboration.

\section{Conclusion}

In this work, we examined how mixed-visual ability work teams manage the accessibility of shared information by conducting diary study with individual interviews and focus groups across seven teams. We documented cases of accessibility-driven transformations and deconstructed how these transformations are triggered, carried out, coordinated, and lead to different consequences. 
Our analysis revealed that transformations are triggered either reactively in the moment or proactively in anticipation of needs, and that coordination can range from explicit negotiation that risks social and professional costs to implicit, relational practices that build trust. We further distinguished between transformations that simplify representations at the cost of functionality and those that enhance them to preserve shared access and utility. 
Across these dynamics, we identified four recurring patterns of teams' action: disposable fixes, transformed becomes the standard, parallel representations, and assembly, each with distinct consequences on labour distribution, collaboration, and equity. 
Our findings point to design opportunities for systems that enable proactive accessibility, support flexible multi-modal representations, encourage beneficial patterns, and balance efficiency with trust-building. 
By analyzing the invisible access labour that enables teamwork, 
we contribute to the understanding of collaborative accessibility in mixed-visual ability work.

\begin{acks}
We thank Shaun Kane for providing thoughtful comments and valuable insights on an earlier draft of this manuscript. 
This research is supported by the University of Victoria, NSERC Canada Graduate Research Scholarship---Doctoral 588775-2024, and NSERC DG 2020-04401. 
\end{acks}

\bibliographystyle{ACM-Reference-Format}
\bibliography{0-bib-main}

\appendix
\section{Summary of Triggers, Actions, and Consequences}
\label{app:sum_TAC}

\aptLtoX{\begin{table}
\centering
\small
\caption{A summary of the triggers, actions, and consequences from the transformation cases.}
\label{tab:TAC_sum}
\begin{tabular}{@{}ll@{}}
\multicolumn{2}{c}{\textbf{(a) Triggers}}\\
\hline
\textbf{Type} & \textbf{Triggers} \\
\hline
\textbf{Work Demand} & Creating and distributing representations (\textit{n}=17) \\
 & Information management or processing (\textit{n}=26) \\
 & Reviewing representations (\textit{n}=22) \\
 & Formal process (e.g., compliance to standard; \textit{n}=19) \\
\hline
\textbf{Representational} & Information is purely visual (\textit{n}=21) \\
\textbf{Incompatibility} & Low-vision challenges (e.g., small font; \textit{n}=11) \\
 & Inconsistent formatting or structure (\textit{n}=10) \\
 & Inoperable elements of representation (\textit{n}=17) \\
\hline
\multicolumn{2}{c}{\textbf{(b) Actions}}\\
\hline
\textbf{Type} & \textbf{Actions} \\
\hline
\textbf{Transformation} & Restructure or reformat (\textit{n}=21) \\
 & Create an alternative representation (\textit{n}=17) \\
 & Create a parallel representation (\textit{n}=12) \\
 & Improve visual access (\textit{n}=5) \\
 & Add alternative access (\textit{n}=18) \\
\hline
\textbf{Coordination} & Mediate (e.g., assistance) (\textit{n}=18) \\
 & Delegate (\textit{n}=10) \\
 & Synchronize (\textit{n}=9) \\
 & Collaborate (\textit{n}=17) \\
 & Advocate (\textit{n}=9) \\
\hline
\multicolumn{2}{c}{\textbf{(c) Consequences}}\\
\hline
\textbf{Type} & \textbf{Consequences} \\
\hline
\textbf{On the Individual} & Enablement and efficiency (\textit{n}=18) \\
 & Increased workload (\textit{n}=13) \\
 & Social cost (\textit{n}=13) \\
 & Professional cost (\textit{n}=8) \\
 & Emotional cost (\textit{n}=9) \\
\hline
\textbf{On the Team} & Increased capacity (\textit{n}=7) \\
 & Increased awareness of others (\textit{n}=14) \\
 & Increased coordination efforts (\textit{n}=5) \\
 & Fragile workflow (\textit{n}=9) \\
\hline
\textbf{On the Representation} & Increased universal access (\textit{n}=14) \\
 & Information silo (\textit{n}=10) \\
 & Decreased fidelity (\textit{n}=9) \\
\hline
\end{tabular}
\end{table}}{
\begin{table}[h]
\centering
\footnotesize
\caption{A summary of the triggers, actions, and consequences from the transformation cases.}
\label{tab:TAC_sum}
\begin{subtable}{\columnwidth}
\centering
\caption{Triggers}%
\label{tab:triggers}
\begin{tabular}{@{}ll@{}}
\toprule
\textbf{Type} & \textbf{Triggers} \\
\midrule
\textbf{Work Demand} & Creating and distributing representations (\textit{n}=17) \\
 & Information management or processing (\textit{n}=26) \\
 & Reviewing representations (\textit{n}=22) \\
 & Formal process (e.g., compliance to standard; \textit{n}=19) \\
\midrule
\textbf{Representational} & Information is purely visual (\textit{n}=21) \\
\textbf{Incompatibility} & Low-vision challenges (e.g., small font; \textit{n}=11) \\
 & Inconsistent formatting or structure (\textit{n}=10) \\
 & Inoperable elements of representation (\textit{n}=17) \\
\bottomrule
\vspace{1pt}
\end{tabular}
\end{subtable}

\begin{subtable}{\columnwidth}
\centering
\caption{Actions}%
\label{tab:actions}
\begin{tabular}{@{}ll@{}}
\toprule
\textbf{Type} & \textbf{Actions} \\
\midrule
\textbf{Transformation} & Restructure or reformat (\textit{n}=21) \\
 & Create an alternative representation (\textit{n}=17) \\
 & Create a parallel representation (\textit{n}=12) \\
 & Improve visual access (\textit{n}=5) \\
 & Add alternative access (\textit{n}=18) \\
\midrule
\textbf{Coordination} & Mediate (e.g., assistance) (\textit{n}=18) \\
 & Delegate (\textit{n}=10) \\
 & Synchronize (\textit{n}=9) \\
 & Collaborate (\textit{n}=17) \\
 & Advocate (\textit{n}=9) \\
\bottomrule
\vspace{1pt}
\end{tabular}
\end{subtable}

\begin{subtable}{\columnwidth}
\centering
\caption{Consequences}%
\label{tab:consequences}
\begin{tabular}{@{}ll@{}}
\toprule
\textbf{Type} & \textbf{Consequences} \\
\midrule
\textbf{On the Individual} & Enablement and efficiency (\textit{n}=18) \\
 & Increased workload (\textit{n}=13) \\
 & Social cost (\textit{n}=13) \\
 & Professional cost (\textit{n}=8) \\
 & Emotional cost (\textit{n}=9) \\
\midrule
\textbf{On the Team} & Increased capacity (\textit{n}=7) \\
 & Increased awareness of others (\textit{n}=14) \\
 & Increased coordination efforts (\textit{n}=5) \\
 & Fragile workflow (\textit{n}=9) \\
\midrule
\textbf{On the Representation} & Increased universal access (\textit{n}=14) \\
 & Information silo (\textit{n}=10) \\
 & Decreased fidelity (\textit{n}=9) \\
\bottomrule
\end{tabular}
\end{subtable}
\end{table}}

\end{document}